\documentclass[10pt]{article}

\usepackage[final]{epsfig}
\usepackage{amsmath}
\usepackage{parskip}

\def\lsim{\mathrel{\rlap{\lower4pt\hbox{\hskip1pt$\sim$}}
    \raise1pt\hbox{$<$}}}         
\def\gsim{\mathrel{\rlap{\lower4pt\hbox{\hskip1pt$\sim$}}
    \raise1pt\hbox{$>$}}}         

\def\overleftrightarrow#1{\vbox{\ialign{##\crcr
    $\leftrightarrow$\crcr
    \noalign{\kern 1pt\nointerlineskip}
    $\hfil\displaystyle{#1}\hfil$\crcr}}}

\begin{document}

\hspace{12cm}{\bf TUM/T39-03-01}\\
\hspace*{12cm}{\bf ECT$^\ast$-03-01}\\

\begin{center}
{\bf Emission of thermal photons and the equilibration time in Heavy-Ion Collisions
\footnote{work supported in part by BMBF, GSI and by the European Commission under
contract HPMT-CT-2001-00370}}
\end{center}
\begin{center}
{Thorsten Renk$^{a}$}

{\small \em $^{a}$ Physik Department, Technische Universit\"{a}t M\"{u}nchen,
D-85747 Garching, GERMANY\\
and ECT$^\ast$, I-38050 Villazzano (Trento), ITALY}

\end{center}
\vspace{0.25 in}

\begin{abstract}
The emission of hard real photons from thermalized expanding
hadronic matter is dominated by the initial high-temperature expansion
phase. Therefore, a measurement of photon emission in ultrarelativistic
heavy-ion collisions provides valuable insights into the 
early conditions realized in such a collision. In particular,
the initial temperature of the expanding fireball or equivalently
the equilibration time of the strongly interacting matter are 
of great interest. An accurate determination of these quantities could
help to answer the question whether or not partonic matter (the quark gluon plasma)
is created in such collisions. In this work, we investigate the emission of real photons
using a model which is based on the thermodynamics of QCD matter
and which has been shown to reproduce a large variety of other observables.
With the fireball evolution fixed beforehand, we are able to
extract limits for the equilibration time by a comparison with
photon emission data measured by WA98.
\end{abstract}

\vspace {0.25 in}

\section{Introduction}
\label{sec_introduction}

In the hot and dense system created in an ultrarelativistic
heavy-ion collision (URHIC), the relevant momentum scales for typical processes
taking place inside the strongly interacting matter drop as a function of proper time
$\tau$: Initially, the relevant scale is set by the incident beam momentum, leading
to hard scattering processes which presumably can be described by perturbative
Quantum Chromodynamics (pQCD). Secondary inelastic scattering processes subsequently
lower the momentum scales due to particle production. At later times, 
equilibration sets in and
typical momenta $p$ are determined by the temperature T of a given 
volume element as $\langle p \rangle = 3T$. As the matter expands, energy stored in random
motion of particles (temperature) is transferred to collective motion
(flow), leading to a descrease of $T$ with $\tau$.
Therefore, by selecting an observable associated with a given momentum
scale, one simultaneously selects a time period in the 
evolution of the system.

For this reason, a measurement of hard real photons is an ideal tool to study
the early moments of the fireball expansion: High momentum photons are not only
sensitive to early proper times, but, being subject to electromagnetic
interactions only, their mean free path in the fireball matter is also much larger than
the spatial dimension of the system, therefore they are capable
of leaving the emission region without significant rescattering.
Therefore, hard photons complement a measurement of low mass, low momentum
dileptons which are dominantly emitted near the kinetic freeze-out point.

Ideally, one would like to use hard photon emission as a thermometer
to determine the initial temperature reached in an URHIC and
use this information to verify the creation of
a quark-gluon plasma (QGP). However,
in reality one has to disentangle thermal contributions to the
photon spectrum from contributions coming from initial hard
scattering processes. The interpretation
of the photon spectrum alone can therefore not be unambiguous.

In this work, we compare a model calculation of photon emission from a fireball
created in an 158 AGeV Pb-Pb collision at SPS with data obtained by the WA98
collaboration  \cite{PhotonData}. 
In a recent paper \cite{Dileptons}, we have developed a fireball
model which is based on information from hadronic observables
and lattice QCD thermodynamics, as manifest in a quasiparticle
picture of the QGP. This model has been shown to successfully
describe low mass dilepton emission \cite{Dileptons} and, within
the framework of statistical hadronization, the measured abundancies
of hadron species \cite{Hadrochemistry}. In the present work, we 
demonstrate that the same model is also capable of describing the
observed photon emission. The fixed setup of the model
also enables us to establish constraints
on the equilibration time $\tau_0$, which entered the
model on an ad-hoc basis so far.

This paper is organized as follows: First, we introduce the photon emission
rate used in the calculation and discuss its interpretation in
the framework of the quasiparticle picture of the QGP which has been used
in the fireball evolution model. In the next section, we summarize
the main properties of the evolution model and discuss constraints
for the initial expansion phase.
Afterwards we present the resulting photon spectrum and demonstrate that,
in agreement with our expectation, hard photons originate dominantly
from the early evolution phase. We investigate the
possibility of using the photon data to set limits on the equilibration
time of the fireball matter and conclude by comparing to the results
obtained by other groups.

\section{The thermal photon emission rate}

\subsection{The QGP contribution}

\begin{figure}[htb]
\begin{center}
\epsfig{file=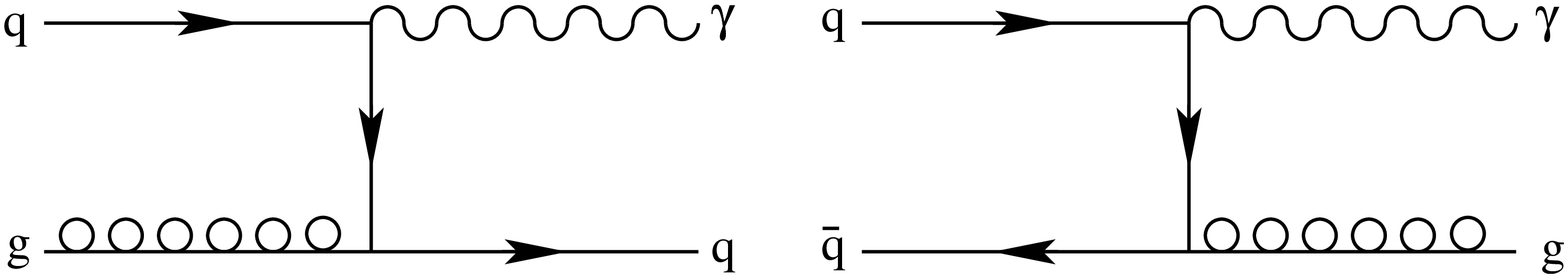, width=8cm}
\epsfig{file=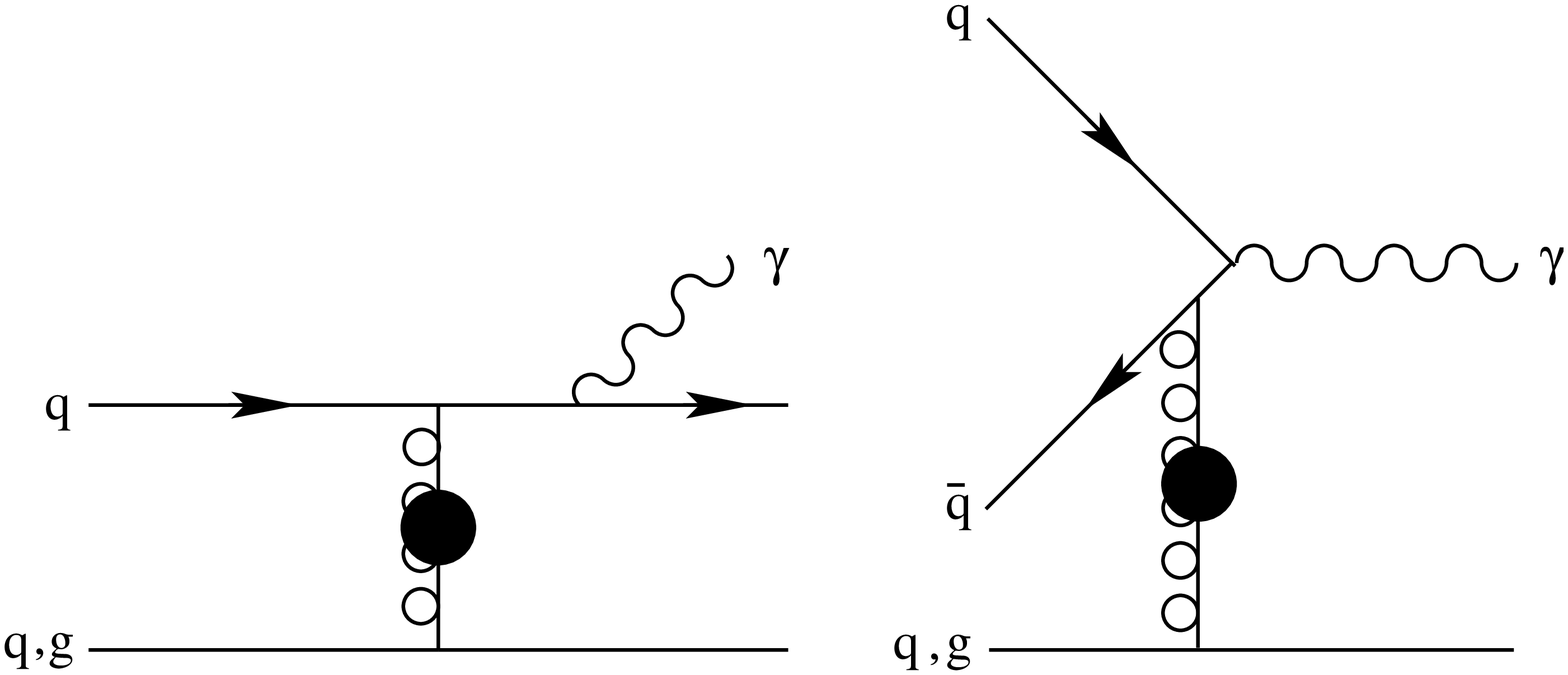, width=7.5cm}
\end{center}
\caption{\label{F-photon-feyn}
Leading order processes for photon production in the QGP (from left to right):
QCD Compton scattering, $q\overline{q}$ annihilation, Bremsstrahlung
and annihilation with scattering (aws).}
\end{figure}

As we expect the dominant contribution to the spectrum of hard
photons to come from a region of high temperatures, we focus
on the QGP rate. To leading order, this
rate can be calculated by evaluating the four diagrams
shown in Fig.\ref{F-photon-feyn}. Here the last two diagrams are
promoted to leading order because of near-collinear singularities.

The complete calculation of the rate to order $\alpha_s$ has been a
very involved task which has been finished only recently
\cite{2-2-Kapusta,2-2-Baier,Aurenche1, Aurenche2, Aurenche3,Complete1, Complete2}.
For the present calculation, we use the parametrization of the rate
given in \cite{Complete2}. There, the rate for photons
of momentum $k$ is written as

\begin{equation}
\label{E-PhotonRates}
\frac{dN}{d^4xd^3k} = \frac{1}{(2\pi)^3}\mathcal{A}(k)\left(\ln(T/m_q(T)) + 
\frac{1}{2} \ln(2E/T) + C_{tot}(E/T) \right),
\end{equation}
with $E=k$ and $m_q^2(T) = 4 \pi \alpha_s T^2/3$ the leading order
large momentum limit of the thermal quark mass.
The leading log coefficient $\mathcal{A}(k)$ reads
\begin{equation}
\mathcal{A}(k) = 2 \alpha N_C \sum_s q_s^2 \frac{m_q^2(T)}{E}
f_D(E).
\end{equation}
Here, the sum runs over active quark flavours and
$q_s$ denotes the fractional quark charges in units of elementary
charge.
The Fermi-Dirac distribution $f_D(E)$ dominates the momentum dependence
of the rate: To a good approximation it decreases exponentially with $E$.
The dependence on the specific photon production process is contained
in the term $C_{tot}(E/T)$: 
\begin{equation}
C_{tot}(E/T) = C_{2 \leftrightarrow 2}(E/T) + C_{brems}(E/T)
+ C_{aws} (E/T).
\end{equation}
All these functions $C(E/T)$ involve non-trivial multidimensional
integrals which can only be solved numerically. In
\cite{Complete2}, parametrizations for the results are
given as
\begin{equation}
C_{2 \leftrightarrow 2}(E/T) \simeq 0.041 (E/T)^{-1}
- 0.3615 + 1.01 \exp[-1.35 E/T]
\end{equation}
and
\begin{equation}
\begin{split}
C_{brems}(E/T)+& C_{aws} (E/T) \simeq \\
\sqrt{1+\frac{1}{6}N_f} 
& \left(
\frac{0.548 \ln(12.28 + 1/(E/T))}{(E/T)^{3/2}} +
\frac{0.133 E/T}{\sqrt{1+ (E/T)/16.27}}
\right).
\end{split}
\end{equation}

\subsection{A quasiparticle interpretation}

In \cite{Dileptons}, we have used a picture of massive, non-interacting
quark and gluon quasiparticles to describe the QGP.
Close to the phase transition these quasiparticles are subject 
to confinement, parametrized
by a universal function $C(T)$ which reduces the number of thermodynamically active
degrees of freedom as 
$n(T) = n_0(T) C(T)$ with $n_0(T) = \int \frac{d^3 p}{(2 \pi)^3} f_{B(D)}(E_p/T)$. 
Here, $f_{B(D)}(E_p/T)$ denotes the Bose (Fermi)
distribution. In \cite{Quasiparticles}, it has been shown that this ansatz
is capable of describing the lattice results for the QCD thermodynamics
extremely well.

In the case of photon emission, we cannot strictly retain this interpretation:
The process $q\overline{q} \rightarrow \gamma$ is kinematically impossible
and all other emission processes would be suppressed by $\alpha_{em}$ for
non-interacting quasiparticles. On the other hand, the fact that
we want to study hard photons implies that at least one of the particles
in the initial state is also hard. But such particles penetrate 
the screening cloud of thermal fluctuations which
is ultimately responsible for the notion of weakly interacting quasiparticles.
Therefore it appears reasonable to allow for interactions of plasma particles
with momenta well above the scale set by the temperature.

The remaining properties of the quasiparticle approach are encoded in the
quasiparticle mass $m(T)$ and the 'confinement factor' $C(T)$.
In the limit of large temperatures, the quasiparticle mass in \cite{Quasiparticles} is chosen
such as to coincide with the result of hard thermal loop (HTL) resummed calculations for the
thermal self-energy. Therefore, inserting massive quarks as degrees of freedom
into the above results would amount to double counting, since those already incorporate 
HTL resummation, at least as long as we consider only temperatures 
above 1.5 $T_C$.

There is no
equivalent of the confinement factor $C(T)$ in the calculations
described in \cite{2-2-Kapusta,2-2-Baier,Aurenche1, Aurenche2, Aurenche3,Complete1, Complete2}.
We can estimate the effect of introducing $C(T)$ as follows:

A typical diagram, say $q\overline{q}$ annihilation, which contributes $R_0$ to the
total emission rate has the structure
$R_0 \sim f_D(E/T)^2 |\mathcal{M}|^2 (1+ f_B(E/T))$, with the thermal quark distributions
$f_D$ in front of the squared matrix element $\mathcal{M}$
corresponding to the process in
vacuum and a Bose enhancement factor for the gluon emitted into
the final state.
The modification of the rate $R$ with respect to the rate $R_0$ in
the presence of $C(T)$ will read $R \sim C(T)^2 f_D(E/T)^2 |\mathcal{M}|^2 (1+ C(T)f_B(E/T))$
which is always larger than $C(T)^3 R_0$. In the case of quarks in the
final state, $C(T)$ leads even to a reduced Pauli blocking (the
final state modification becomes $(1-f_D(E/T) C(T))$ which is larger
in the presence of confinement).

On the other hand, as mentioned above, at least one of the incoming particles
has to have a large momentum. Such 
high-momentum particles are not subject to confinement along with the
bulk of the matter. One would expect them to hadronize well
outside the thermalized region, leading to $C(1) = 1$ at one of the
incoming legs in the diagram.

Therefore, $C(T)^3$ can be taken as a conservative estimate of the effect
of the confinement factor on the emission rate. If it can be shown that
hard photon emission is dominated by a region where the temperature is
so large that $C(T) \approx 1$, the above expression for the emission rate
can be used to approximately describe photon emission from a system
of quasiparticles also. This is also the region where we expect the
mass of quasiparticles to be given by the HTL result.
We will verify this property a posteriori.

Clearly, the prescription outlined here has to be regarded as an
approximation till a more detailed version of a quasiparticle description
of the QGP incorporating confinement is available.

\subsection{The hadronic contribution}

As the temperatures in the hadronic evolution phase of the fireball
are lower than in the QGP phase, we expect the hadronic contributions
to the emission of hard photons to be small. Therefore, we will not
discuss this contribution in great detail.

Vector mesons play an important role for the emission
of photons from a hot hadronic gas. The first calculation
of such processes has been performed in \cite{Kapusta-Eff}
in the framework of an effective Lagrangian.
It has been found that the dominant processes
are pion annihilation, $\pi^+\pi^- \rightarrow \rho \gamma$,
'Compton scattering', $\pi^\pm \rho \rightarrow \pi^\pm \gamma$
and $\rho$ decay, $\rho \rightarrow \pi^+\pi^- \gamma$.

Several more refined approaches have been made since then
(for an overview, see \cite{PhotonReview}).
In the following, we will use a parametrization of the
rate from a hot hadronic gas taken from \cite{HHG} which
is given as
\begin{equation}
E \frac{dN}{d^4xd^3k} [\text{fm$^{-4}$ GeV$^{-2}$}]
\simeq 4.8 T^{2.15} \exp[-1/(1.35 ET)^{0.77}] \exp[-E/T].
\end{equation}

\subsection{The integrated rate}

In order to compare to the experimentally measured photon spectrum \cite{PhotonData},
we have to integrate Eq.~(\ref{E-PhotonRates}) over
the space-time evolution of the fireball,
\begin{equation}
\label{E-PhotonIntegratedRate}
\left. \frac{dN}{d^2 k_t dy}\right|_{y=0}=
\frac{ \pi}{\Delta y}
\int d \tau R^2(\tau) \int_{z_{min}(\tau)}^{z_{max}(\tau)} 
\negthickspace \negthickspace dz
\int_{k_{min}(y(z))}^{k_{max}(y(z))} d k_z \frac{dN}{d^4xd^3k}.
\end{equation}
In this expression, $R(\tau)$ stands for the radial expansion of the
fireball, $\Delta y$ denotes the rapidity interval covered by the
detector, $y(z)$ is the rapidity of a volume element
at position $z$ and the limits of the $k_z$ integration come from
the fact that a photon emitted at the (boosted) edge of the fireball
has to have a longitudinal momentum in a certain range in
order to be detected in the rapidity window of the experiment.
In this expression, we have assumed spatial homogeneity and a cylindrical fireball.

\section{The fireball evolution model}

In this section, we briefly outline the general framework of the
fireball evolution model. The model is described in greater detail
in \cite{Dileptons} and \cite{Thesis}.

The underlying assumption of the model is that the strongly
interacting matter produced in the initial collision reaches thermal
equilibrium at a timescale $\tau_0 \approx 1$ fm/c.
For simplicity, we assume spatial homogeneity of all thermodynamic
parameters throughout a 3-volume at a given proper time. The evolution
dynamics is then modelled by calculating the thermodynamic response to
a volume expansion that is consistent with
measured hadronic momentum spectra at freeze-out.

The volume itself is taken to
be cylindrically symmetric around the beam (z-)axis. In order to
account for collective flow, we boost individual volume elements inside
the fireball volume with velocities depending on their position.
As flow velocities in longitudinal direction turn out to be close to the
speed of light, we have to include the effects of time dilatation.
On the other hand, we can neglect the additional time dilatation caused 
by transverse motion, since
typically $v_\perp \ll v_z$. The thermodynamically relevant volume is
then given by the collection of volume elements corresponding to the
same proper time $\tau$. In order to characterize the volume expansion
within the given framework, we need first of all the expansion velocity
in longitudinal direction as it appears in the center of mass frame. 
For the position of the front of the cylinder, we make the ansatz
\begin{equation}
z(t) = v_0\, t + c_z
\int_{t_0}^{t} \!\! d t' \!\int_{t_0}^{t'} \!\!\!d t''
\ \frac{p(t'')}{\epsilon(t'')}\,,
\end{equation}
where $c_z$ is a free parameter.
The time $t$ starts running at $t_0$ such that $z_0 = v_0 \,t_0$ is
the initial longitudinal extension with $v_0$ the initial longitudinal 
expansion velocity. The longitudinal position $z(t)$ and $t$ itself define 
a proper time curve $\tau = \sqrt{t^2 - z^2(t)}$. Solving for 
$\tilde t = t(\tau)$ one can construct $\tilde z(\tau) = z(\tilde t)$.
Then the position of the fireball front $z(t)$ in the center of mass frame 
can be translated into the longitudinal extension $L(\tau$) of the cylinder
on the curve of constant proper time $\tau$. One obtains
\begin{equation}
L(\tau) = 2\int_0^{\tilde z(\tau)}\!\!ds\ \sqrt{1 + \frac{s}{\sqrt{s^2 + \tau^2}}}.
\end{equation}
At the same proper time we can define the transverse flow velocity and construct
the transverse extension of the cylinder as a circle of radius
\begin{equation}
R(\tau) = R_0 + c_\perp 
\int_{\tau_0}^{\tau} \!\! d \tau' \!\int_{\tau_0}^{\tau'} \!\!\!d \tau''
\ \frac{p(\tau'')}{\epsilon(\tau'')}\,
\end{equation}
where $R_0$ corresponds to the initial overlap radius, and $c_\perp$
is a free parameter. With the values of $L$ and $R$ obtained at a given
proper time we can parametrize the 3-dimensional volume as 
\begin{equation}
V(\tau) = \pi R^2(\tau) \  L(\tau)\,.
\end{equation}

In principle the model is fully constrained by fixing the
freeze-out time $t_f$ (at the fireball front), 
the proper time $\tau_f$ and the two constants $c_\perp$ and $c_z$,
such that the measured hadronic observables are reproduced. In practice this is achieved
only at SPS energy, where the freeze-out analysis \cite{FREEZE-OUT}
allows a complete characterization on the basis of hadronic
$dN/dy$ and $m_t$-spectra and HBT radii. For the 5\% most central Pb-Pb collisions 
one requires
\begin{equation}
R(\tau_f) = R_f,\quad v_\perp(\tau_f) = v_\perp^f,\quad v_z(\tau_f) = v_z^f
\quad \text{and} \quad T(\tau_f) = T_f
\end{equation}
while making a trial ansatz for the ratio $p/\epsilon$. Note that the
proportionality of the acceleration to this ratio (which is reminescent of the
behaviour of the speed of sound) allows a soft
point in the EoS to influence and delay the volume expansion.

Assuming entropy conservation during the expansion phase, we fix
the entropy per baryon from the number of produced
particles per unit rapidity. 
Calculating the number of participant baryons as 
\begin{equation}
N_p(b) = \int d^2 s\ n_p({\bf b},{\bf s}) \label{participants} 
=  \int d^2 s\ \,T_A(s) \left[1-\exp(-\sigma^{in}_{NN}T_B(|{\bf b} -{\bf s}|)\right]+(T_A 
\leftrightarrow T_B)\,,  
\end{equation}
we find the total entropy $S_0$. The entropy density at a given proper time
is then determined by $s=S_0/V(\tau)$. Using the EoS given by the 
quasiparticle approach \cite{Quasiparticles}, thereby providing the link with lattice QCD, we find 
$T(s(\tau))$ and also $p(\tau)$ and $\epsilon(\tau)$, which in the next 
iteration replace the trial ansatz in the volume parametrization.
Finally we arrive at a thermodynamically self-consistent 
model for the fireball which is by construction able to describe the
hadronic momentum spectra at freeze-out.

In order to find the fireball evolution relevant for the photon emission
measured for the 10\% most central collisions, we follow the procedure outlined in
\cite{Dileptons}. In brief, we consider an effective system which starts
out with a reduced number of participants and hence reduced total entropy
content. Neglecting azimuthal asymmetries,
we keep parametrizing the expanding system as a cylinder with  reduced
initial radius. Assuming that the freeze-out temperature is approximately
unchanged for more peripheral collisions, we determine a reduced proper
evolution time and modify the geometrical freeze-out radius and the
transverse flow velocity accordingly. However, going from
the 5\% to the 10\% most central collisions, we find differences in the early
evolution phases on the level of a few percent only.

Thus, the fireball evolution is completely constrained by hadronic observables. 
In \cite{Dileptons}, it has been shown that this scenario is able to
describe the measured spectrum of low mass dileptons, and
in \cite{Hadrochemistry} it has been demonstrated that under the assumption
of statistical hadronization at the phase transition temperature $T_C$,
the measured multiplicities of hadron species can be reproduced.
None of these quantities is, however, sensitive to the detailed choice
of the equilibration time $\tau_0$. Therefore, we have only considered
the 'canonical' choice $\tau_0 =$ 1 fm/c so far. The calculation of photon emission
within the present framework provides the opportunity to test this assumption
and to limit the choice of $\tau_0$.

\section{Results}

\begin{figure}[!htb]
\begin{center}
\epsfig{file=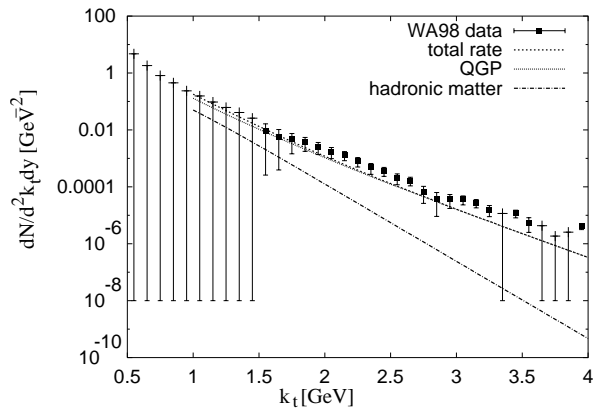, width=8cm}
\end{center}
\caption[Spectrum of thermal photons for 10\% most central
Pb-Pb collisions at SPS, 158 AGeV Pb-Pb collisions, model calculation]
{\label{F-photon-rates}
Thermal photon spectrum for 10\% most central
Pb-Pb collisions at SPS, 158 AGeV Pb-Pb collisions, shown are
calculated rate (total, contribution from QGP and hadronic gas) and
experimental data \cite{PhotonData}.}
\end{figure}

The result of the evaluation of Eq.(\ref{E-PhotonIntegratedRate}) with the
fireball evolution model described in the previous section is shown in
Fig.~\ref{F-photon-rates}.

The overall agreement with the data is remarkably good.
Above 2 GeV, the calculation underestimates the
data somewhat, leaving room for a contribution of prompt photons from initial hard processes
of about the same magnitude as the thermal yield. Note that
the spectrum is almost completely saturated by the QGP contribution ---
for $k_t > 3 $ GeV, the hadronic contribution is almost two orders of magnitude down.
This can in essence be traced back to the strong
temperature dependence of the emission rate normalization
and justifies the approximate treatment of the hadronic contribution
a posteriori.

\begin{figure}[!htb]
\begin{center}
\epsfig{file=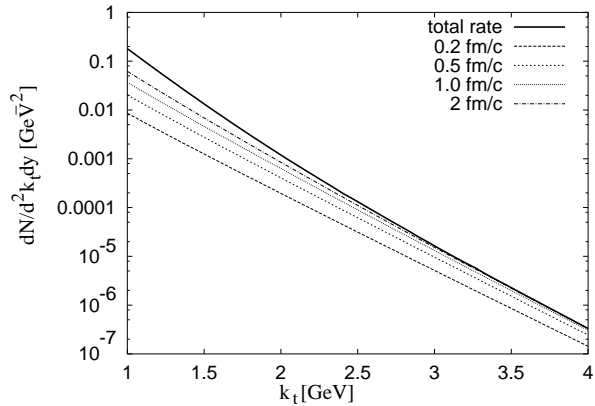, width=8cm}
\end{center}
\caption{\label{F-photon-taudep}
The total photon emission spectrum and the integrated rate at proper times
$\tau = 0.2, 0.5, 1.0$ and 2 fm/c of the fireball evolution.}
\end{figure}

In order to study the importance of the initial, high temperature phase in 
more detail, we present the time evolution of the spectrum in Fig.~\ref{F-photon-taudep}.

One observes that the large $k_t$ region is almost exclusively dominated
by the first fm/c of evolution proper time, whereas the yield
in the low $k_t$ region is not yet saturated after 2 fm/c.
However, the temperature associated with these evolution times is
always larger than  250 MeV, leading to $C(T) > 0.9$ \cite{Quasiparticles} and
$C(T)^3 \approx 0.7$. This justifies neglecting the effect of quasiparticle
properties on the rate a posteriori: There is a 30\% uncertainty introduced
into the calculation, but this is comparable with other intrinsic
uncertainties, such as the detailed choice of the numerical value of $\alpha_S(T)$ in
this temperature regime. 

\begin{figure}[!htb]
\begin{center}
\epsfig{file=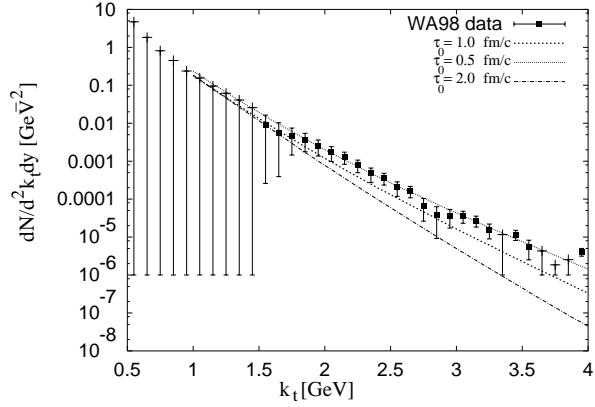, width=8cm}
\end{center}
\caption[The thermal photon emission spectrum for different
choices of the equilibration time $\tau_0$ as compared to experimental data]{\label{F-photon-tau0}The thermal photon emission spectrum for different
choices of the equilibration time $\tau_0$ as compared to experimental data
\cite{PhotonData}.}
\end{figure}

The high $k_t$ tail of the spectrum is potentially capable of providing
information about the initial temperature reached immediately after
equilibration. This capability is seriously limited
in practice, however, by the need to assess an unknown contribution of
prompt photons, which may be large in this region.
Bearing this uncertainty in mind, we can nevertheless pursue this
idea further in Fig.~\ref{F-photon-tau0} where we
investigate the sensitivity of the result to the equilibration time
$\tau_0$ of the fireball.

We find that the low $k_t$ region of the spectrum is hardly
affected by different choices for the equilibration time,
while for larger $k_t$ one is increasingly sensitive to short evolution 
timescales.
An equilibration time of 0.5 fm/c corresponding to an initial
temperature of 370 MeV leads to a good description of the data
without the inclusion of any prompt photon contribution. 
On the other hand, a rather slow equilibration corresponding to $\tau_0 = 2$ fm/c
and an initial temperature of 260 MeV requires a sizeable contribution
from prompt photons.

Without any reference to prompt photons, we are therefore able
to fix $\tau = 0.5$ fm/c as the lower bound for the equilibration
time: Shorter timescales would lead to thermal photon
emission overshooting the data.

\begin{figure}[htb]
\begin{center}
\epsfig{file=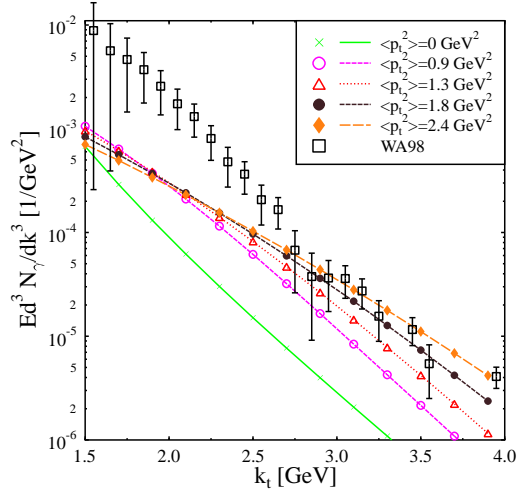, width=8cm}
\end{center}
\caption[Prompt photon production rates in Pb-Pb collisions for different values
of average parton intrinsic transverse momentum $\langle p_t^2 \rangle$
as compared to experimental data]
{\label{F-k_tdep}
Prompt photon production in Pb-Pb collisions 
as a function of the photon transverse momentum $k_t$ for different values
of average parton intrinsic transverse momentum $\langle p_t^2 \rangle$ \cite{Lars}
as compared to experimental data \cite{PhotonData} (figure adapted 
from \cite{Lars}).}
\end{figure}

If we want to find an upper limit for the equilibration time, we have to address the
issue of prompt photon emission.
For this purpose, we use the results of two different works \cite{Lars, Wong} to
illustrate the possible range of predictions dependent on the average value
of $p_t$, an 'intrinsic' transverse momentum scale
which is introduced as a phenomenological parameter
to account for non-perturbative effects. (Fig.~\ref{F-k_tdep}).

One observes that prompt photon production is
able to explain the data above 3 GeV for $\langle p_t^2 \rangle  \sim 1.8$ GeV$^2$,
but no choice of $\langle p_t^2 \rangle$ leads to a description of the
data below 3 GeV. 
Thus, 
there is a momentum region between 2 and 2.5 GeV where the data are only weakly
affected by a prompt photon contribution.

On the other hand, the present calculation of thermal photon emission indicates
sensitivity to the equilibration time in this region
(see Fig.~\ref{F-photon-tau0}). One finds that an equilibration time of
2 fm/c is unable to describe the data even in the presence
of a sizeable prompt photon contribution, hence we may regard this as
the upper limit for the equilibration time given the results of \cite{Lars}.

If we use alternatively the results of \cite{Wong} to estimate the contribution
of prompt photons, the situation is somewhat less clear, though the
main conclusions remain valid. In the momentum region near 2 GeV, the result
obtained for the prompt photon contribution in \cite{Wong} is about
50\% of the thermal emission spectrum obtained with $\tau_0 = 1$ fm/c in the present
calculation,
which in turn is already on the lower end of the error bars of the data. Therefore, 
the prompt photon
contribution alone is unable to explain the spectrum in this momentum region,
and a thermal contribution with  $\tau_0 \approx 2$ fm/c is needed to
describe the data near $k_t = 2$ GeV.
(In \cite{Wong}, it is stated that the prompt photon contribution is
of the same size as the thermal emission spectrum corresponding to
a fireball with initial temperature of 300 MeV, which is not true
in the present calculation. However, this statement is
derived using Bjorken hydrodynamics, whereas the present model
incorporates longitudinal acceleration of the fireball matter which leads 
to less cooling initially and hence to a prolonged QGP phase.)

However, there are several uncertainties in the calculation. We have already
discussed the complications introduced by the quasiparticle picture and
the detailed choice of $\alpha_S$. An additional potential problem in
the calculation is that the rate has been calculated to order $\alpha_S$,
but $\alpha_S$ is not a small quantity in the relevant temperatur region,
so there might be sizeable higher order corrections. In view of these uncertainties 
and the less urgent need for an addition al contribution to prompt photons
if the results of \cite{Wong} are taken,
it is possibly better to estimate 0.5 fm/c $< \tau_0 <$ 3 fm/c for the equilibration time.

In principle, one might try to construct a scenario with a large
equilibration time, using the most optimistic estimate for the prompt
photon contribution. Below temperatures $\sim 200$ MeV, the approximations made
to calculate the photon emission from the QGP phase break down, however
qualitatively we expect the confinement factor $C(T)$ to strongly reduce the
emission rate. Likely candidates for photon yield between 2 and 2.5 GeV
are therefore only the hadronic evolution phase and the pre-equilibrium phase.

In order for the hadronic phase to contribute significantly, either the emission
rate or the four-volume of emitting matter needs to be increased significantly.
The four-volume of fireball matter in the hadronic phase, however, is tightly
constrained by the measured amount of dilepton radiation, which has been
discussed in the present framework in \cite{Dileptons}. An increase of the
emissivity of a hadronic gas, on the other hand, would very likely
be accompanied by a change in the number of active degrees of freedom,
which presumably are driven by in-medium mass reductions of resonances
as argued in \cite{MassShifts}. 
A strong increase of active degrees of freedom, however, 
is not in agreement with a statistical hadronization
analysis done in \cite{Hadrochemistry}. Even an overall mass
reduction of 10\% (excluding the pseudoscalars) is
not compatible with the observed hadron ratios. Therefore, within the present
framework, the hadronic phase is not likely to give a large
contribution to the photon spectrum.

It remains the question of the yield from the pre-equilibrium phase.
No rigorous scenario leading to thermal equilibrium for SPS conditions
has been developed so far, however, several aspects of the pre-equilibrium
dynamics have already been investigated.

In \cite{Preeq-Alam}, the kinetic equilibration of different quark flavours was investigated under
the assumption that gluons come to an early equilibrium and constitute a heat bath in which
quark motion takes place. An early non-equilibrium distribution of quarks 
would of course directly influence the photon emission spectrum. Somewhat related is the question
of chemical equilibration of quarks. Here, the hot-glue scenario \cite{HotGlue}
has been suggested where an initial undersaturation of the
quark densities with respect to the thermal equilibrium densities is assumed, i.e. almost
all of the entropy of the system is carried by the gluons, leading to a drastically increased
initial temperature. 

The findings of \cite{Preeq-Alam} suggest, albeit for RHIC
conditions, that the typical timescale for the kinetic equilibration of light
quark flavours is of order $\sim 1$ fm/c. This timescale is roughly in line with
the assumptions made in the present work. However, it leaves the question if the photon
emission signal is affected if one starts with a suitable out-of-equilibrium
initial quark distribution.

Regarding the hot-glue scenario, note that the drastically increased temperature 
of the partonic matter would mostly affect the high momentum tail of observed photons
and therefore leave the momentum region below 2.5 GeV less affected. This is
especially true since also the overall normalization of the emission rate is reduced
with respect to equilibrium conditions due to the undersaturation of the quark densities.

A different approach to pre-equilibrium dynamics has been taken in
\cite{Preeq-Kaempfer}. Here, transverse momentum dependence of dilepton
emission has been calculated in a kinetic framework and calculations for different
initital parton distributions have been tested. While the approach is very interesting,
it is hard to directly estimate its possible influence on photon emission
within the present framework.

In \cite{Preeq-Bass}, an investigation of non-equilibrium photon emission
has been carried out using a parton-cascade model (PCM).
Here, the essential findings were that only a very dilute partonic medium
is created in the collision. Photon emission from this medium was shown
to explain the data above 3 GeV when integrated up to the hadronization
point, but below 3 GeV, the photon spectrum from the partonic phase
falls below the data. It is difficult to relate these findings
directly to the present approach, since no equilibrium phase in
either partonic or hadronic phase is described in the PCM. However, we may
take this as an indication that pre-equilibrium dynamics is most likely
to strongly affect the high momentum region of the photon spectrum only where we find
considerable uncertainties with regard to the question of intrinsic $p_t$ anyway.

In the present work, no attempt has been made to calculate a contribution
to the photon spectrum from pre-equilibrium matter. It is, however, unlikely that a long-lasting
pre-equilibrium phase is characterized by a strong photon emission rate,
since strong photon emission indicates frequent interaction processes in
the medium which in turn would lead to fast equilibration. Furthermore, it is
plausible that a pre-equilibrium contribution mainly affects the (uncertain)
high momentum region of the spectrum, as it is characterized by hard momentum
scales. Nevertheless, all results discussed in this paper are subject to some uncertainties
resulting from the poor knowledge of pre-equilibrium dynamics.

\section{Comparison to other works}

Several other scenarios have been investigated by different
authors in order to explain the photon spectrum measured by WA98.
In \cite{Srivastava}, a hydrodynamical evolution model has been used.
The favoured scenario found in this work uses a very short equilibration time
of 0.2 fm/c, corresponding to an (average) initial temperature  of 335 MeV.
The contribution of thermal photons is about 50\% of the total yield, the rest
is prompt photon contribution.

While the result for the equilibration time is clearly very different from
the findings of the present work, several other results are similar,
among them the weak sensitivity to the phase transition temperature $T_C$
(indicating the dominance of early emission phases), the favoured large
initial temperature and the dominance
of the QGP over the hadronic signal. It remains to explain the remarkable
difference in the conclusions about the equilibration time.

Note that both in  \cite{Srivastava} and the present work
large initial temperatures $T_i \gsim 300$ MeV are needed to give a sizeable
thermal photon emission in the momentum range in question. The choice of
$\tau_0 = 0.2$ fm/c seems mainly driven by the need to create such
large $T_i$. There are, however, two important differences
between our model and the one in \cite{Srivastava} which lead naturally
to large initial temperatures for equilibration times above 0.5 fm/c
in our approach.

First, we employ an EoS as based on lattice results, which leads to 
a temperature increase of about 30\% for a given entropy density as compared
to a bag model EoS. Second, the temperatures quoted in \cite{Srivastava}
are obtained using the Bjorken estimate \cite{Bjorken}. Our fireball evolution, however,
incorporates significant longitudinal acceleration of matter, which in essence
leads to a peaked initial distribution of energy density at central
rapidities and hence to significantly larger initial temperatures.

In \cite{Huovinen}, a number of scenarios with different EoS and
initial state have been investigated within a hydrodynamical description. 
The reference scenario described there uses
a bag model EoS for the QGP phase with a transition temperature of
180 MeV and an initial state which leads to a peak initial temperature
of $T_i^{max} =325$ MeV and an average initial temperature of
$\overline{T}(z=0) = $ 255 MeV. There is clearly a discrepancy between
the average initial temperature in \cite{Huovinen} and the present work.
Due to the different space-time expansion patterns of the hot matter
in the (averaged) present calculation and a hydrodynamics evolution, this
issue could possibly best be clarified by comparing the amount
of four-volume corresponding to a given temperature instead of comparing
the average at a given $\tau$.
At the moment however, this has to be regarded as an open question.

In addition, there is a larger contribution to the photon
spectrum of matter with temperatures
below 200 MeV observed in \cite{Huovinen}  (about 40\% of the total yield)as compared to our
model. 
This is presumably caused the different choice of the
EoS, which in the bag model case leads to faster cooling and to a
long-lasting mixed phase, in essence reducing the weight of
contributions from high temperatures and enhancing the low-
temperature yield.

\section{Summary}

A measurement of hard real photon emission provides a
good opportunity to study the early evolution of a fireball.
Due to the temperature dependence of the photon emission rate,
the QGP phase is expected to dominate over the contribution from
the hadronic phase. We have used the leading order photon
emission rate from the QGP, along with estimates of the impact
of our phenomenological quasiparticle picture on this rate,
in a simple model for the fireball evolution to calculate the
resulting photon spectrum. As this model was fixed beforehand, we
have not introduced any new free parameters.

For the 'standard choice' of the equilibration time $\tau_0$, the
model has been able to give a good description of the data.
Nevertheless, we have tried variations of this quantity in order
to work out constraints. With the help of an estimate for the
contribution of photons from initial, hard processes,
we found for the equilibration time 0.5 fm/c $< \tau_0 <$ 3 fm/c.
This would imply that a QGP phase
must be present in the evolution, at least within the present model.

Overall, our picture of the spacetime evolution of the fireball
finds now support from both the low momentum (late time) and
the high momentum (early time) region of the evolution.
More precise future measurements and calculations
can be expected to tighten the constraints on the equilibration time.

\section*{Acknowledgements}

I would like to thank W. Weise, S.S.~R\"{a}s\"{a}nen, R.A. Schneider and
A. Polleri for interesting and stimulating discussions and helpful
comments.


\begin{thebibliography}{99}

\bibitem{PhotonData}
M.~M.~Aggarwal {\it et al.}  (WA98 Collaboration),
nucl-ex/0006007.

\bibitem{Dileptons}
T.~Renk, R.~A.~Schneider and W.~Weise,
Phys.\ Rev.\ C {\bf 66} (2002) 014902.

\bibitem{Hadrochemistry}
T.~Renk,
hep-ph/0210307.

\bibitem{2-2-Kapusta}
J.~Kapusta, P.~Lichard and D.~Seibert,
Phys.\ Rev.\  {\bf D 44} (1991) 2774.

\bibitem{2-2-Baier}
R.~Baier, H.~Nakkagawa, A.~Niegawa and K.~Redlich,
Z.\ Phys.\ {\bf C 53} (1992) 433.


\bibitem{Aurenche1}
P.~Aurenche, F.~Gelis, R.~Kobes and H.~Zaraket,
Phys.\ Rev.\  {\bf D 58} (1998) 085003.

\bibitem{Aurenche2}
P.~Aurenche, F.~Gelis and H.~Zaraket,
Phys.\ Rev.\ D {\bf 61} (2000) 116001.

\bibitem{Aurenche3}
P.~Aurenche, F.~Gelis and H.~Zaraket,
Phys.\ Rev.\ D {\bf 62} (2000) 096012.

\bibitem{Complete1}
P.~Arnold, G.~D.~Moore and L.~G.~Yaffe,
JHEP {\bf 0111} (2001) 057.

\bibitem{Complete2}
P.~Arnold, G.~D.~Moore and L.~G.~Yaffe,
JHEP {\bf 0112} (2001) 009.


\bibitem{Quasiparticles}
R.~A.~Schneider and W.~Weise,
Phys.\ Rev.\ C {\bf 64} (2001) 055201.

\bibitem{Kapusta-Eff}
J.~I.~Kapusta, P.~Lichard and D.~Seibert,
Phys.\ Rev.\ {\bf D 44} (1991) 2774.

\bibitem{PhotonReview}
T.~Peitzmann and M.~H.~Thoma,
Phys.\ Rept.\  {\bf 364} (2002) 175.

\bibitem{HHG}
F.~D.~Steffen and M.~Thoma,
Phys.\ Lett.\ {\bf B 510} 2001 98.


\bibitem{Thesis}
T.~Renk, PhD Thesis.

\bibitem{FREEZE-OUT}
B.~Tomasik, U.~A.~Wiedemann and U.~W.~Heinz, nucl-th/9907096.





\bibitem{Lars}
A.~Dumitru, L.~Frankfurt, L.~Gerland, H.~Stocker and M.~Strikman,
Phys.\ Rev.\ {\bf C 64} (2001) 054909.

\bibitem{Wong}
C.~Y.~Wong and H.~Wang,
Phys.\ Rev.\ C {\bf 58} (1998) 376.



\bibitem{MassShifts}
J.~e.~Alam, P.~Roy, S.~Sarkar and B.~Sinha,
nucl-th/0106038.

\bibitem{Preeq-Alam}
J.~Alam, B.~Sinha and S.~Raha,
Phys.\ Rev.\ Lett.\  {\bf 73} (1994) 1895.

\bibitem{HotGlue}
E.~V.~Shuryak, Phys. Rev. Lett {\bf 68} (1992) 3270,
E.~V.~Shuryak and L.~Xiong, Phys. Rev. Lett {\bf 70}
(1993) 2241.

\bibitem{Preeq-Kaempfer}
B.~Kampfer and O.~P.~Pavlenko,
Phys.\ Rev.\ C {\bf 49} (1994) 2716.

\bibitem{Preeq-Bass}
S.~A.~Bass, B.~Muller and D.~K.~Srivastava,
Phys.\ Rev.\ C {\bf 66} (2002) 061902.


\bibitem{Srivastava}
D.~K.~Srivastava and B.~Sinha,
Phys.\ Rev.\ C {\bf 64} (2001) 034902.

\bibitem{Bjorken}
J.~D.~Bjorken, Phys.\ Rev.\ {\bf D 27}, (1983) 140.

\bibitem{Huovinen}
P.~Huovinen, P.~V.~Ruuskanen and S.~S.~Rasanen,
Phys.\ Lett.\ B {\bf 535} (2002) 109.


\end{thebibliography}
\end{document}